\documentclass{ae}[times]  
\usepackage{graphicx}
\usepackage{txfonts}
\usepackage{hyperref}
\begin{document} 

   \title{{\em \Large{Very Important\\Letter to the Editor}}\vspace{0.6cm}\\ When Tails Tell Tales}
\titlerunning{When Tails Tell Tales}
\authorrunning{Henri}

   \author{Henri M.J. Boffin
          \inst{1}\fnmsep\thanks{As for all his papers, this author is sole responsible for its content, which does not represent in any way or another, not even when seen through a telescope, the views of his employer, real or supposed.}
                 }

   \institute{$^1$ Extraterrestrial Institute for the Advancement of Earth (EIAE),
            Secret place, Planet Earth, Solar System  
                           }

   \date{Received March 29, 2023; accepted March 30, 2023}

  \abstract
   {The enigmatic open clusters serve as a constant reminder of the mysteries of the universe, helping to confront astronomical theories. Unknown to many, these clusters often possess tails with inappropriate labels, serving as the tell-tale signs of their historical journey. But unlike typical tails, these extensions can either precede or follow the body, yet they consistently unfold a cosmic mystery to be solved. 
   I present a succinct survey of this subject matter, detailing the intrepid efforts of astronomers who have dared to challenge our knowledge about these creatures, and offer a novel proposal for their nomenclature, while not disregarding the philosophical ramifications.
 
   \vspace{0.25cm}}
   \maketitle
\begin{flushright} {\it In the open sky, among the twinkling stars,\\
There dance the clusters, with their tails afar.\\
\^O clusters bright and bold, your tails stretched wide,\\
Guiding you through the galaxy with pride.
 }\\  
\end{flushright}

\section{A community of stars}
In the vast expanse of the heavens, there exists a celestial wonder that has captured the imagination of astronomers and stargazers alike. This wonder is known as an open cluster -- a gathering of stars that twinkle in the night sky like diamonds on a velvet cloth. Like a bustling village, an open cluster is a community of stars that come together to form a dazzling spectacle. Each star has its own unique character and personality, but they all share a common bond as members of the same cluster. It is a sight that fills the heart with wonder and the mind with curiosity, as we seek to unravel the mysteries of these celestial communities and the secrets they hold within their glowing hearts.

This close-knit community shares a common origin and moves together through the galaxy. Stars in an open cluster interact with each other and are affected by their environment. The cluster's position and motion within the Milky Way can also have an impact on its stars, just as the location and resources available to a village can affect the lives of its inhabitants. Similarly, just as people in a village may come and go, stars in an open cluster can be added or lost over time as a result of various interactions and processes. It is this buzzing activity, this relentless motion that the astronomer wants to capture.

To identify the members of an open cluster is a task that requires the patience of a saint and the eye of an eagle. It is a delicate dance of observation and deduction, where the slightest misstep can lead one astray. But through careful study and analysis, the members of the cluster can be brought into focus, each one a shining example of the wonders of the universe. With the members identified, the celestial sleuth can embark into its journey of discovery, delve deeply into the intricate details of its population and confronting observations with theoretical models, up to the point to even coming closer to unlocking the elixir of youth, the key to the fountain of life itself \citep{B21}.

\begin{figure*}
    \begin{center}
        \includegraphics[width=9cm]{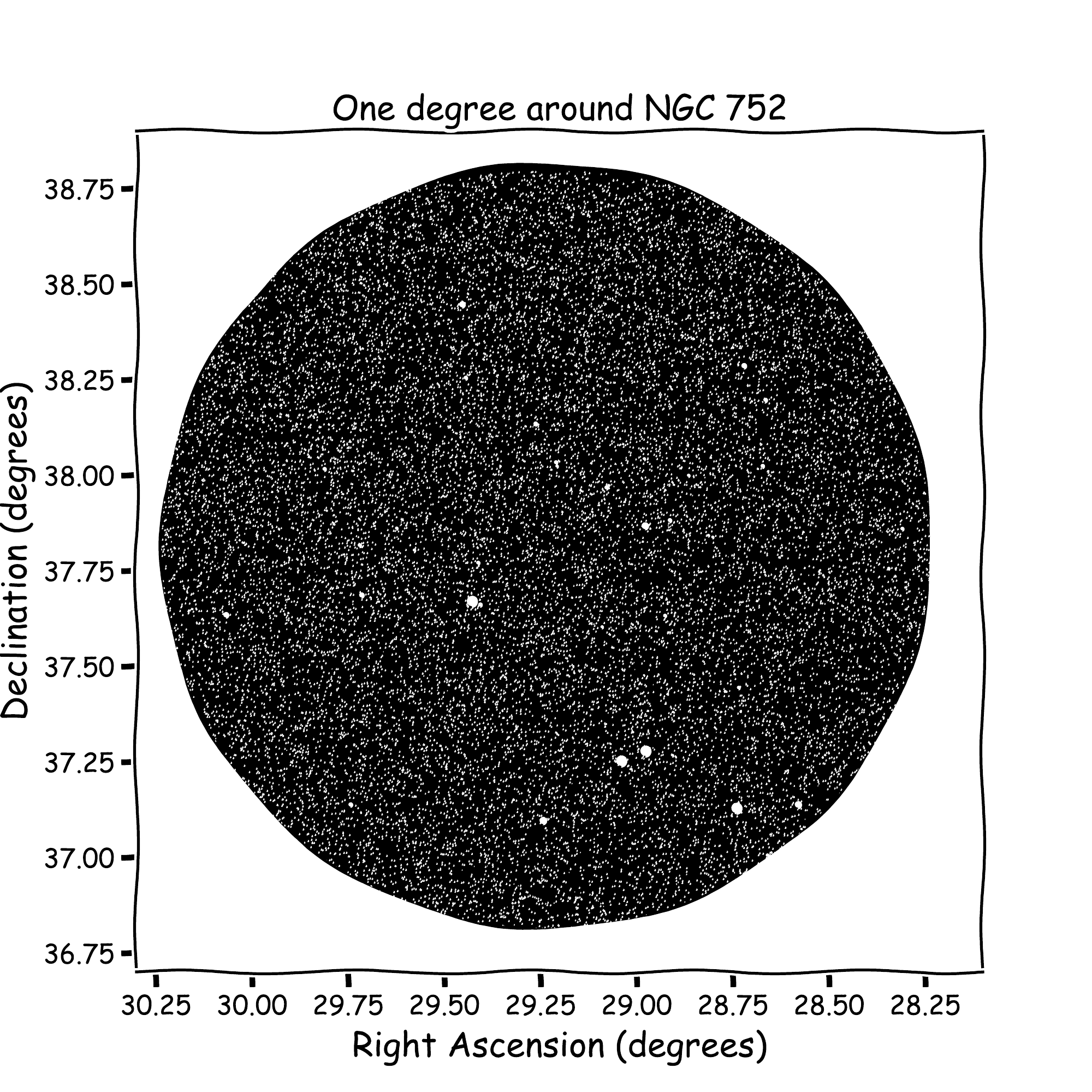}\includegraphics[width=9cm]{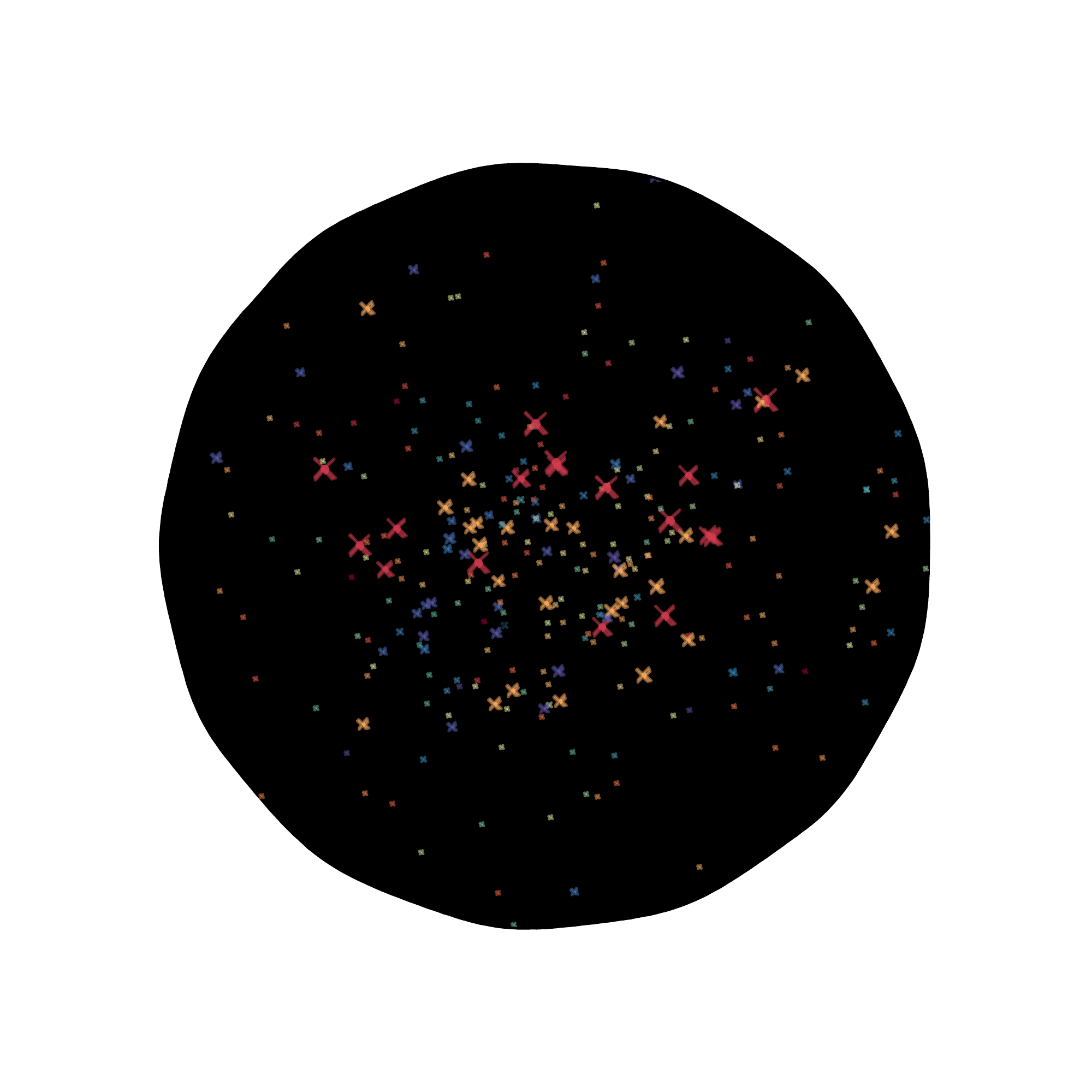}
    \caption{(left) Behold, a vista of 1 degree's radius encircling the heart of the open cluster NGC 752, whereupon the majority of stars detected by Gaia are revealed. Yet, amidst this boundless expanse of twinkling gems, how can one hope to discern the true members of this cluster? Fear not, for the answer is revealed in the right panel, wherein the clever fusion of machine learning techniques in a quintic space allows us to unveil these members. From a prodigious total of over 30,000 stars in the left panel, we have plucked a mere 295 cluster members, whose hues are selected to correspond with their varying temperatures. Take heed, however, that the radiance of these two panels differ, as the brightest stars depicted in the left panel are not affiliated with the cluster in any way. The right panel attempts to simulate the appearance of stars through a telescope with terrible optics.}
    \label{fig:h}
    	\end{center}
\end{figure*}

But how is this possible, my dear reader? How do we manage to discern the members of such a cluster from the vast expanse of stars that surround it? How does one sift through the stars like grains of sand, and separate the wheat from the chaff?
We must be patient and persistent, casting our nets into the boundless sea of stars until we find what we seek (Fig.~\ref{fig:h}). 

Back in the days of yore, when astronomy was not yet the leisurely indoor pastime it is now, the identification of the members of an open cluster was a most arduous task. It was a bit like looking for a needle in a haystack -- except the needle is invisible and the haystack is the vast expanse of the cosmos. You would have to don your finest astrophysical gear and find yourself a nice, cozy spot under the stars. Then, you'll need to peer through your telescope with the sharpness of a hawk's eyes and the patience of a saint, trying to find stars that seemingly moved in unison.

But fear not, my friend, to have to resort to such tedious endeavour, for the power of Gaia\footnote{\url{https://www.cosmos.esa.int/gaia}} is now here to help you! First, take a deep breath and calm your mind, because you're going to need all your wits about you.
Next, head on over to the Gaia archive and download the data for your target cluster. Look for the parallax and proper motion measurements, and use them to calculate the cluster's position and motion in the sky. This will help you determine where the cluster is located relative to the background stars. But the fun doesn't stop there! Gaia also provides information about each star's brightness and color. By studying these properties, astronomers can further refine their selection of cluster members. So there you have it  -- a lighthearted guide to detecting the members of an open cluster using data from Gaia. With a bit of patience, persistence, and a dash of humor, you'll be able to uncover the secrets of these celestial objects in no time!

Mayhap the above discourse does sound a mite too arduous to apprehend, thus permit me to present it in a humbler guise, drawing upon an analogy that may be better within your ken. Imagine you're at a party and you don't know anyone. You want to figure out who's actually part of the party and who's just wandering around aimlessly. So, with anxiety, you start by looking at their name tags (parallax), which tells you how far away they are from you. Then, you look at their place in the room (positions) to see if they're standing in a group of people or just lurking in a corner. But that's not enough. You also want to know if they're actually participating in the party or just hanging out, so you look at how they're moving (proper motions) to see if they're dancing, talking, or just standing still. All of this information together helps you figure out who's part of the party and who's just crashing it. And that's basically what scientists do when they use Gaia data to find members of an open cluster. But let's be honest, it's a lot less fun than a party.

With this simile, no doubt did you, dear reader, realise that such analysis requires a quintic space that captures not just the location of stars, but also their movements and parallax -- a space much more appropriate than our familiar, but so limited, 3-dimensional universe. To understand this concept, one might think of the classic novel ``Flatland’’ \cite{A84}, where the inhabitants of a two-dimensional world are visited by a three-dimensional being. To them, this being appears as a series of mysterious shapes that change in size and form as it moves through their world. Similarly, the 5-dimensional space of Gaia data allows us to see the stars in a new way, not just as static points of light in the sky, but as dynamic entities that move and interact with one another. By analysing the parallax and proper motion of stars, we can determine their membership in open clusters. In passing, it may be of interest to note that the same technique was also used quite remarkably to predict the exact date of the doom of Humankind \citep{B14}.

\section{Moving beyond the cluster}
The daring astronomers seeing how well they were able to distinguish the members of an open cluster -- {\it de facto} mapping a village -- started to envision extending this study to look out for any stars that seem to be straying from the pack. Once they've found a few of these rogue stars, they'll need to connect the dots, so to speak, and trace their path through the galaxy. And voila! quickly enough, they found the tails of open clusters -- the tell-tale signs of their past, leading or trailing behind like a cosmic detective story.

The open cluster is a village, scattered across the vast cosmic expanse, with each star shining as a unique house. But it is not alone; it is accompanied by the phantom arms of its tidal tails, reaching out into the emptiness like spectral appendages. These arms whisper secrets of the cluster's past, of its growth and decline, of the forces that shape it. And in the midst of it all, we stand, mere observers, attempting to comprehend the beauty and mystery of this cosmic tableau.

The tails of an open cluster can be compared to the roads leading away from a small village. Just as the village is a center of community life, the open cluster is a center of gravitational attraction for its member stars. And just as the roads branching out from the village allow for the spreading of houses into the surrounding countryside, the tidal tails of the cluster allow its stars to spread out into the Milky Way.
Yet, as with any village that loses its inhabitants to the wider world, the dissolution of the cluster through its tails brings a sense of loss and nostalgia for what once was. The stars, once bound together in the tight-knit community of the cluster, now wander alone in the vast expanse of space. It is a bittersweet fate, one that speaks to the transience of all things and the impermanence of our place in the universe.

Verily, this is the destiny of open clusters, for as soon as they come into existence, they feel the tidal forces of the mighty Milky Way, much like the gravitational pull of the Moon on the oceans of Earth that causes the tides to rise and fall. Stars at the edge of the cluster start to drift away. They are pulled by an invisible force, and their trajectories are forever altered. The other stars watch in confusion and horror as their companions are torn away from them.  As time passes, more and more stars follow suit. They are dragged out of the cluster by the same invisible force. But as they drift away, they leave something behind: a long, wispy trail that stretches out behind them like a ghostly tail. 

The stars were sad to see their companions go, and realized that they too would eventually meet the same fate.
But they were not afraid. They would dissolve into the field, just like their friends before them, and they would become something new, something different, something beautiful.

The tidal forces that act on an open cluster are like the unseen hand of fate that guides us towards our destiny. They are like the winds of change that sweep through our lives, shaping our path and influencing our journey.
The tidal forces can be both gentle and fierce, subtle and strong, like the gentle caress of a breeze or the ferocity of a storm. They shape the very fabric of the universe, creating patterns that are both intricate and mesmerizing.
In many ways, the tidal forces are a reflection of our own lives. They remind us that we are all connected, that our actions have consequences, and that we are all part of a larger whole. 
In the end, the stars of an open cluster dissolve through their tidal tails into the field to populate the Milky Way, just as we all dissolve into the fabric of the universe when our time has come. It is a humbling reminder that we are all but a small part of something much larger, a reminder to cherish every moment and to live our lives to the fullest.

Perhaps you can’t make head or tail of it? Perhaps you are lost in space? Or find yourself quite befuddled? Allow me to elucidate this matter in plain language for the less poetic, and more voracious, among the readers. 
Tidal forces acting on an open cluster are like the irresistible urge to eat a bag of potato chips - you know you shouldn't, but you just can't help it! The cluster's outer stars get pulled away like chips from the bag, slowly dissolving and dispersing into the galaxy's vast snack aisle. And just like with a bag of chips, once you start snacking on those outer stars, it's hard to stop until they're all gone. So, beware the gravitational munchies!

\section{The story of a tadpole}
One might say that the narrative of the tails of an open cluster bears resemblance to that of Benjamin Button, who began his life as an old man and aged backwards, ending as a newborn baby. But, unlike Button, a cluster starts its life as a frog to end as a tadpole. And, indeed, clusters have  been compared to tadpoles, with the bright cluster core being the head and the dimmer, more spread-out outer regions being the tail. As the tails of the clusters dissolve, they lose their original shape and eventually disappear, leaving behind only the stars that once formed them. And in the same way as the tail dissolves, so too will the cluster eventually disperse into the vast expanse of the universe, its stars continuing on their own individual journeys, forever changed by their time spent in the embrace of the cluster.

In terms of size, the tidal tails of a cluster can vary widely, depending on a number of factors, such as the mass of the cluster, the speed of its motion through space, and the strength of the gravitational forces acting upon it.
Some tidal tails can be relatively short and stubby, while others can stretch out for many light-years, trailing behind the cluster like a comet's tail.

These tails can reveal much about the dynamics and history of the cluster, as well as about the invisible forces acting upon it. One may even hope that these tails will, one day, help us understand whether there is really a need for this elusive substance -- the modern ether of astrophysics -- that is called dark matter. Indeed, the study of these tails can offer us insights into the very nature of gravity itself. For it is through the analysis of these celestial formations that we can begin to unravel the complex interplay of forces that shape the movements of the celestial bodies that populate our universe.

\begin{figure*}
    \begin{center}
        \includegraphics[width=18cm]{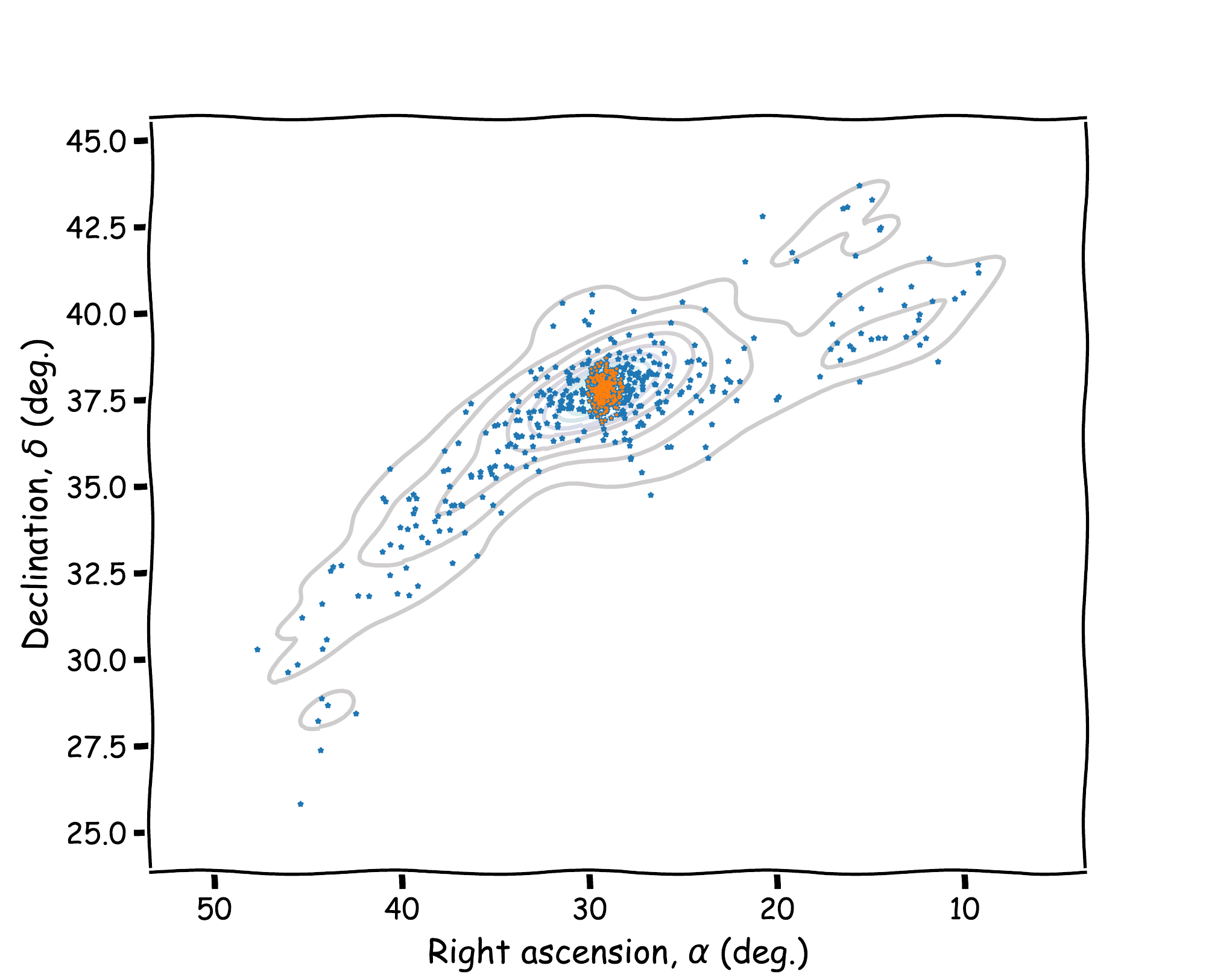}
    \caption{The amazing tails -- or shall we accept to call them arms? -- of the open cluster NGC 752 are unveiled through the selection of their members via the convergent point method. The blue dots represent all members, with the orange dots indicating those belonging to the core of the cluster, as presented in Fig.~1. The arms stretch for approximately 850 light-years (tip to tip) and exhibit evident asymmetry. The number of stars in the outer regions of the arms is equivalent to those in the core.  }
    \label{fig:ads}
    	\end{center}
\end{figure*}

\section{The ice cream truck}
If the majority of tails found are rather modest in size, very quickly, the most elongated appendices were found attached to some remarkable open clusters -- some being several hundreds of light-years long \citep{2021A&A...647A.137J,2022MNRAS.514.3579B}! The reason for this is that some astronomers are real gluttons for punishment, unafraid to explore the most daring of the techniques and get their hands dirty. Instead of simply use the proper motions of the stars to look for a group of stars that are all hanging out together in the vastness of space, they resorted to the convergent point method, which involves some serious mathematical calculations that would make your head spin faster than a top. 

To use the convergent point method is like looking for a group of people all walking towards the same ice cream truck. If you stand in the right spot, you'll see them converging on the truck from all directions. Now, imagine those people are stars, and the ice cream truck is the center of an open cluster. The convergent point is where all their paths intersect, and that's how you find the former cluster's members. Of course, it's not always as easy as finding an ice cream truck. Sometimes the stars are scattered all over the place, like a group of kids chasing after different toys in a toy store. But with a little detective work and some clever calculations, you can still track them down using the convergent point method.
And while both methods are used to establish the membership of stars in open clusters, the convergent point method has proven to be far more accurate than the use of proper motions for extended structures. This is due to its ability to take into account the curvature of the celestial sphere and the relative motions of the stars. Proper motions, on the other hand, make the audacious assumption that stars move in straight lines at a fixed velocity, a premise which is not always veracious. It is akin to attempting to capture a bunch of rambunctious tots prancing around a playground with a straight ruler, while the convergent point method is more akin to utilising a GPS to trace their movements and foretell their final destination. Put differently, it is the difference between endeavouring to hit a moving target with a blindfold and having the aid of a high-tech tracking device. I have no doubt about which method you would use, dear reader, provided that you are not intimidated by the ostensible complexity of the mathematics involved.

The astronomers who followed this daring path thus found most extended tails. Henceforth, if the previously discovered were similar in shape to tadpoles, the newly found ones are sometimes compared to the mighty appendices of a chameleon (Fig.~\ref{fig:ads}) --  long and winding snakes, slithering and twisting through the blackness of space, that are made up of stars that had strayed from their clusters, wandering off on their own paths through the galaxy. They are a symbol of the stars' independence and freedom, a reminder that even in the vastness of space, there was still room for exploration and adventure.

Inaccurate is the comparison to a chameleon, however, for the tails of certain open clusters expand up to hundred times their size. After a study whose length and thoroughness defy all imagination, I came to the conclusion, dear reader, that only one animal, with its elongated tail, bears resemblance: the Marsupilami! 

Created in 1952 by the Belgian cartoonist Andr\'e Franquin, this fictional creature was first featured as a supporting character in the illustrious comic book series, {\it Spirou et Fantasio}, which was widely read and enjoyed by the masses. Franquin, being a master of his art, imbued the Marsupilami with a certain charm and charisma that captured the hearts of many, and it was not long before the creature became a star in its own right. Indeed, the Marsupilami's popularity grew so much that it was given its own spin-off series, where it could shine in all its glory, and where its adventures and misadventures could be fully explored. 

This marvellous creature is unlike any other, being highly intelligent and nimble with a prehensile tail that can stretch to wondrous lengths. Many a naturalist have puzzled over this creature, wondering what sort of evolutionary advantage such a lengthy tail might provide. Some speculate that it is used for balance or to help the animal navigate through dense underbrush. Others suggest that it is a weapon, used to stun prey or defend against predators.
But in truth, the Marsupilami's tail is a marvel of versatility. It allows the creature to move with incredible grace and agility, twisting and turning as it leaps through the treetops. And when the Marsupilami needs to communicate with its fellows, it can wave its tail in a complex series of gestures and signals, conveying a wealth of information in the blink of an eye.
While the Marsupilami draw inspiration from creatures of the natural world such as the opossum and the lemur, its specific traits and talents are entirely the stuff of fantasy.
It is my uttermost conviction, nevertheless, that Franquin, with his prophetic eye, has fashioned the Marsupilami in the very image of the tails of open clusters. 

\section{Let's correct this improper wording}
Until now, those of you who have managed to keep their eyes open must be feeling an irresistible urge to scold the author in the most cantankerous manner: ``Cease this chatter about tails! A tail is the appendage that trails behind a body, whereas your clusters possess conspicuous extensions on either side, hence you are driven to employ oxymorons such as `leading tails’!’’ 

I can only concur with this discerning observation, dear reader, but will likely further disappoint you as unfortunately this isn’t the sole use of improper wording in the field of open clusters. Indeed, it is rather common usage to speak for example of the corona of an open cluster to refer to the outer region of the cluster that contains stars that are gravitationally weakly bound to the main body of the cluster. These stars are more susceptible to being stripped away by external gravitational forces, such as those from passing stars or the tidal forces of the Milky Way, and may eventually escape the cluster altogether. The corona can extend out to several times the radius of the main body of the cluster and can contain a significant fraction of the cluster's total mass.
 
What a misnomer! What a curious choice of words, for such a term could not be further from the truth. Verily, it appears that the names we give things are as fickle and capricious as the winds that blow over the moors. Unlike the crown atop a monarch's head, th corona is made of the less massive and less luminous stars in the cluster, as these are the ones more susceptible to be thrown out, and they are thus the less to attract the attention of all who gaze upon them. And given the pandemic that Humanity is still going through, we should stay away from words like ``corona’’. Why not call it a ``coma’’ as, similarly to the diffuse, fuzzy cloud around the nucleus of a comet that is caused by the sublimation of its ices, we are talking about the evaporated stars escaping from the cluster? 
 
Hence, instead of erroneously discussing tails, one ought to discourse about colossal arms, extending across the cosmos, which, instead of seizing new stars to augment their assemblage, actually relinquish them. Just as a mighty jungle tree sends out its branches in search of sunlight and nourishment, these open clusters reach out with their glowing appendages. And just like a tree's branches, these arms are covered in a verdant array of colors, from pale blue to fiery red, like the leaves that rustle in the breeze.

\newpage
\section{Famous last words}

\begin{flushright} {\it And so, the open clusters disperse,\\
Their beauty and wonder, a fleeting verse.\\
A reminder of the ephemeral nature of all things,\\
Of the impermanence of even the brightest kings.\\

Thus, open clusters, like life itself,\\
Are fleeting and ethereal, like an elf.\\
They sparkle, they shine, and they fade away,\\
A fleeting beauty that's not meant to stay.
}\\  
\end{flushright}

The tidal arms of open clusters are like a long scarf which unrolls under the tidal force of the Milky Way, a gentle yet persistent force that shapes their destiny. Each strand of the scarf tells a story of stars that once belonged to the cluster, but were gradually pulled away by the tide until they became part of the wider galaxy. And yet, the scarf keeps unrolling, as more stars leave the cluster. In this unequal dance between the cluster and the galaxy, the scarf is the only witness to the passing of time. It stretches across the sky like a ghostly veil, a reminder of the stars that were, all bound together by the same force that shapes the universe.

As I gaze at the scarf of tidal arms, I cannot help but wonder: is this the fate of all things in the cosmos, to be swept away by the tides of time and space? Or is there a hidden order that governs the dance of the stars, a secret pattern that eludes our grasp?
I may never know the answer, but one thing is certain: the scarf of tidal arms will keep unrolling, a silent witness to the mystery of the universe. And I, a mere mortal, can only stand in awe and wonder at the beauty and complexity of it all.

\vspace{0.5cm}
\begin{acknowledgements}
As for the  other  pieces in this series, this work was done outside of working hours, when the first author was bothered by his annoying colleagues, A. Wake, W.H.Y. Can't, I. Sleep. The author made use of the OpenAI ChatGPT 3.5, which was a true test of the author's fortitude, as it felt like attempting to extract coherence from an inebriated individual who went back into childhood and had lost touch with all sense of reality -- in the words of the developers, a system that may ``make up facts or `hallucinate' outputs''!

This work has made use of data from the European Space Agency (ESA) mission
{\it Gaia} (\url{https://www.cosmos.esa.int/gaia}), processed by the {\it Gaia}
Data Processing and Analysis Consortium (DPAC,
\url{https://www.cosmos.esa.int/web/gaia/dpac/consortium}). Funding for the DPAC
has been provided by national institutions, in particular the institutions
participating in the {\it Gaia} Multilateral Agreement.
This research has made use of NASA’s Astrophysics Data System Bibliographic Services.
Matplotlib \citep{2007CSE.....9...90H} and its {\it xkcd} environment are gratefully acknowledged.
\end{acknowledgements}

\end{document}